\documentclass{aa}
\usepackage{graphicx,times}
\newlength{\fsize}
\newcommand{\mk}{}
\newcommand{\mkb}{}

\def\Real{{\rm I\mathchoice{\kern-0.70mm}{\kern-0.70mm}{\kern-0.65mm}%
  {\kern-0.50mm}R}}  
  \def\bx#1{\leavevmode\thinspace\hbox{\vrule\vtop{\vbox{\hrule\kern1pt
  \hbox{\vphantom{\tt/}\thinspace{\bf#1}\thinspace}}
  \kern1pt\hrule}\vrule}\thinspace}

\def\be{\begin{equation}} \def\ee{\end{equation}}

\begin{document}

\title{The stability of poloidal magnetic fields in rotating stars}

\author{J. Braithwaite\thanks{\emph{Current address:}
Canadian Institute for Theoretical Astrophysics, 60 St.~George St., Toronto M5S 3H8, Canada \tt jon@cita.utoronto.ca}}
\institute{Max-Planck-Institut f\"ur Astrophysik, Karl-Schwarzschild-Stra{\ss}e 1,
Postfach 1317,  D--85741 Garching, Germany
}
\authorrunning{J. Braithwaite}
\titlerunning{The stability of poloidal magnetic fields in rotating stars}
\date{~~}

\abstract{
The stability of large-scale magnetic fields in rotating stars is
explored, using 3D numerical hydrodynamics to follow the evolution of
an initial poloidal field. It is found that the field is subject to
an instability, located initially on the magnetic equator, whereby
the gas is displaced in a direction parallel to the magnetic axis.
If the magnetic axis is parallel to the rotation axis, the rotation
does not affect the initial linear growth of the instability, but does restrict the growth of the
instability outside of the equatorial zone. The magnetic energy decays on a
timescale which is a function of the Alfv\'en crossing time and the
rotation speed, but short compared to any evolutionary timescale. No
evidence is found for a possible stable end state to evolve from an
initial axisymmetric poloidal field. The field of an oblique rotator is similarly unstable, in
both cases regardless of the rotation speed.

}

\maketitle

\section{Introduction}
\label{sec:intro}

The question of the stability of large-scale fields in stars has accompanied the historical development of
magnetohydrodynamics (MHD) right from the beginning. Over the same
period, large-scale magnetic fields have been detected in
a variety of stars -- since their discovery in Ap stars
(\cite{Babcock:1947}) they have also been observed (or their presence
inferred) in white dwarfs, neutron stars, upper-main-sequence stars,
central stars of planetary nebulae, etc. (e.g. \cite{Kempetal:1970}, \cite{Angeletal:1974},
\cite{Henrichsetal:2003}, \cite{Jordanetal:2005}). It seems
implausible that these fields can be regenerated by a
convective dynamo. {\mk This is because, in contrast to solar-type main sequence stars, these stars have either no significant convective zone, in the case of white dwarfs and neutron stars, or just a small convective core, in the case of upper-main-sequence stars. {\mkb Any field produced in this convective core would struggle to rise to the surface on a sensible timescale (see, e.g. \cite{Moss:2001}). There is also the lack of one hallmark of dynamo activity -- the correlation between rotation rate and magnetic field strength found in solar-type and some other stars (\cite{BorandLan:1980}, \cite{Mathysetal:1997}).} For this} reason there has been great interest in finding some
magnetic field configuration which is stable on a sufficiently long
timescale and which can therefore persist without recourse to a
continuous generative mechanism.

By means of the following non-rigorous argument, one can see that it
should be possible to construct a magnetic field in equilibrium,
i.e. where the Lorentz, gravity and pressure forces cancel
everywhere. Because of the zero-divergence constraint, the magnetic
field, and therefore the Lorentz force, has two degrees of
freedom. Luckily, we have two degrees of freedom in making adjustments
to the thermodynamic state of the star, e.g. to the
temperature and pressure fields. In this way, the net force on every
fluid element can be brought to zero.

\begin{figure}
\includegraphics[width=1.0\hsize,angle=0]{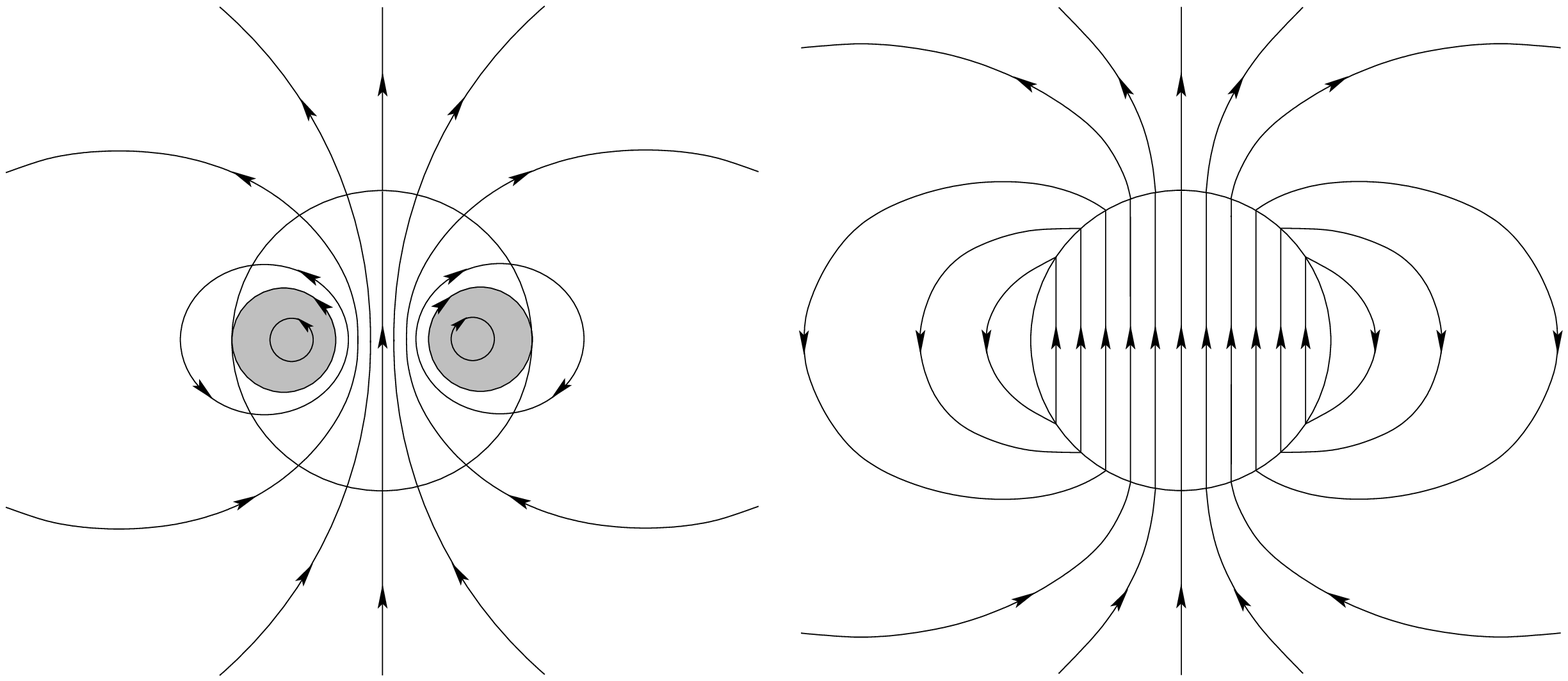}
\caption{Two magnetic field configurations. On the left, the
mixed poloidal-toroidal field (shading in diagram represents toroidal
field) shown to be stable in a non-rotating
star (\cite{BraandSpr:2004}). On the right, a purely poloidal field. All purely poloidal fields are known to be
unstable in a non-rotating star.}
\label{fig:stable-and-ums}
\end{figure}

The next question is that of the stability of an equilibrium to an
arbitrary perturbation. {\mk Tayler (1973) examined purely toroidal fields (i.e. having only the azimuthal component $B_\phi$)
in non-rotating stars, finding necessary and sufficient conditions for stability. It is generally supposed that these stability conditions would be impossible to satisfy at every point in the star. The resulting instability, }which is global in the
azimuthal direction, will grow on the timescale given by the time
taken for an Alfv\'en wave to travel across the star ($\sim 10$ years
in a $1M_\odot$, $1R_\odot$ star with {\mk a} field of $1$ kilogauss). Some
properties of this instability have now been analysed numerically by
Braithwaite (2006). It has also been shown that any purely poloidal
field (i.e. having only components $B_r$ and $B_\theta$ in
spherical polars; one
example of a poloidal field is illustrated on the right-hand-side of
fig.~\ref{fig:stable-and-ums}) is also unstable in a non-rotating star
(Markey \& Tayler 1973, 1974, \cite{Wright:1973},
\cite{BraandSpr:2006}). {\mk At least in a non-rotating setup,} a stable field had therefore to be of a mixed
toroidal-poloidal form, and Wright (1973) suggested that a toroidal
field could indeed be stabilised by adding a poloidal field of
comparable strength.

{\mk These previous studies described above worked by taking {\mkb various} field configurations and examining their stability, by either analytic or, in the case of \cite{BraandSpr:2006}, numerical means. A different approach is to take an arbitrary magnetic field, which one assumes will be not only not in stable equilibrium but not in any equilibrium, and to follow its evolution in time, to see if it eventually finds a stable equilibrium on its own. A study of this kind has now been done:
following the evolution of an arbitrary initial field using numerical
MHD, a stable equilibrium was} found (\cite{BraandSpr:2004},
\cite{BraandNor:2006}). This stable field is of a mixed
poloidal-toroidal twisted-torus shape. Moreover, this was the {\it
only} stable equilibrium found, starting from various different
initial magnetic fields. It is illustrated qualitatively on the left-hand-side of
fig.~\ref{fig:stable-and-ums}.

It seems likely therefore that in a {\it non-rotating} star, there is only
one possible stable magnetic field configuration. The question of what
is stable and unstable in a {\it rotating} star, however, has
historically received less attention, largely because of the
increased complexity of the problem. A useful method of demonstrating 
stability in a non-rotating system is to show that for all arbitrary
perturbations, the energy of the system increases. If even only one perturbation
can be found for which the energy decreases, the system is
unstable. Unfortunately, this method does not work in a rotating
system. To understand why, imagine trying to push a ball off the
top of a hill. The potential energy of the ball falls as it moves
further from the summit, and one could conclude that the equilibrium at the top of the hill is an unstable one. However, a Coriolis force will cause the ball
to move in a curve and eventually to come back to where it started (on an idealised hill without friction).

Frieman \& Rotenberg (1960) demonstrated that solid-body rotation would not have
a significant effect on hydromagnetic instabilities unless the
rotation velocity was at least comparable to the hydromagnetic
velocity. This is the case for some, but not all, magnetic stars. {\mk[Indeed, within each group, including neutron stars, white dwarfs and upper main sequence stars, there are some slow rotators and some fast rotators; for instance, some neutron stars (magnetars) have an Alfv\'en crossing time and rotation timescale ($\Omega^{-1}$) of $0.1$s and $1$s respectively, while a typical radio pulsar might have corresponding timescales of $100$s and $0.1$s.]}
Pitts \& Tayler (1985) looked at these faster rotating stars, concluding
that instabilities in toroidal fields {\mk are unlikely to be} stabilised entirely, rather
the growth rate is merely reduced by a factor roughly equal to the ratio
of rotation and hydromagnetic velocities, and
that the instabilities in poloidal fields are unlikely to be suppressed, unless (perhaps)
the magnetic axis is inclined at a large angle to the rotation axis.

More recently, Geppert \& Rheinhardt (2006) have used numerical MHD to
look at the stability of poloidal fields in rotating stars, finding
that if the star is rotating quickly enough, and if the field is
roughly aligned to the rotation axis, the instability is suppressed. This is
at odds with the results presented here; this difference is discussed in section~\ref{sec:concs}.

In this study, I use numerical methods to investigate the stability
of poloidal fields in rotating stars. The paper is organised as
follows. In the next section, I review the form of the instability
expected to be present in non-rotating stars. The model used in the investigation is outlined in
section~\ref{sec:methods}, as well as some of the technical and computational
details. I present the results in section~\ref{sec:results}, and
finish with a summary of the conclusions and a discussion of their
implications in section~\ref{sec:concs}.

\section{The form of the instability in poloidal fields}
\label{sec:instability}

In the results obtained so far with analytic methods, a distinction
has been made between poloidal fields where some of the field lines
are closed within the star and poloidal fields where all field lines
pass through the stellar surface. However, the results (in the
non-rotating case) have been the same: an instability whose growth
rate is comparable to the Alfv\'en crossing time. 

The cause of the instability (in at least the latter type of poloidal field) can be understood intuitively by means of
the following line of reasoning (\cite{FloandRud:1977}). Imagine the
star as being composed of two solid halves, each containing a
frozen-in magnetic field, and free to rotate about a common axis. The
star now resembles two parallel bar magnets. The bar magnets will of
course rotate until the north pole of one is next to the south pole of
the other, and the star can do exactly the same. Thinking of
this in energy terms, the magnetic energy stored in the star is
unchanged but the energy in the atmosphere has fallen, providing
energy to drive the instability. One difference between the bar magnets and
the star is that while the bar magnets are only unstable to one mode
of the instability (i.e. the azimuthal wavenumber $m=2$), the
star is subject also to instability at higher wavenumbers. 

\begin{figure}
\includegraphics[width=1.0\hsize,angle=0]{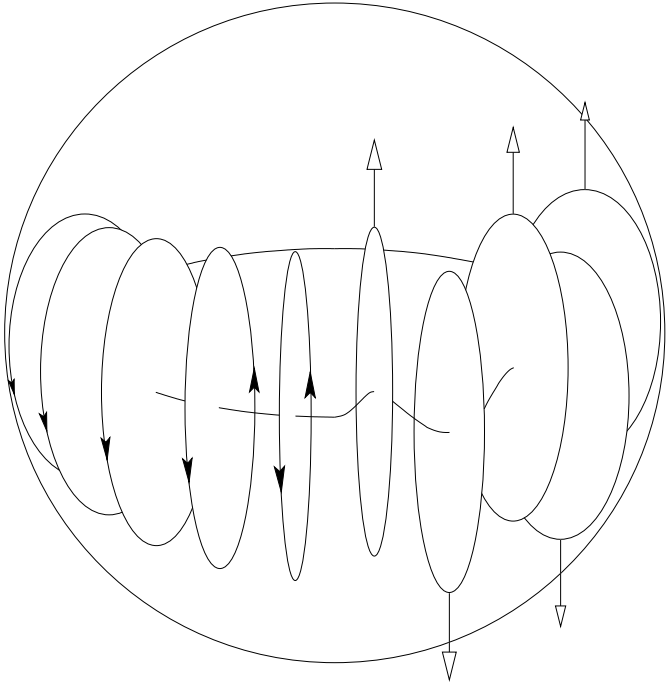}
\caption{Loops of poloidal magnetic field in a star. The
left-hand-side of the
star is in equilibrium. On the right, the form of the instability is
shown -- displacements are {\mk essentially} parallel to the magnetic axis and
perpendicular to gravity. For clarity, a line is drawn connecting the
centres of each field loop.}
\label{fig:form-of-inst}
\end{figure}

Even when some of the field lines are closed within the star, the bar
magnets analogy is still of some use, since rotating one half of the star by
$180^\circ$ will still result in a decrease in magnetic energy outside
of the star. {\mkb Theoretically is it possible that {\it all} field lines are closed within the star, (it seems unlikely that a star born out of a molecular cloud could be completely cut off from the cloud's field, but in some stars there could plausibly be a mechanism on the stellar surface to bury the magnetic field entirely from view as is sometimes described in models of accreting neutron stars,) and in this case (as in the other two cases just described)
it} is helpful to think in terms of the
magnetic field loops exerting pressure on each other and there being a
tendency for them to slip out of the equilibrium position (or in other words, be pushed out by
pressure from the neighbours). Since the
movement of these loops is restricted in the radial direction by the
stable stratification of the star, they move in a direction parallel
to the magnetic axis. This is illustrated in
fig.~\ref{fig:form-of-inst}.

{\mk It has been shown that a system of two bar magnets can indeed be stabilised by rotation (see \cite{Jonesetal:1997}). However, there is reason to believe that the analogy between bar magnets and stars does not hold here, because a star is fluid and has more degrees of freedom. More precisely, the magnetic field loops described in the last paragraph move parallel to the magnetic axis, and if this coincides with the rotation axis, the Coriolis force will have no effect,} {\mkb at least in the limit of small displacements.}

\section{The numerical model}
\label{sec:methods}

The numerical model used is identical to that described in
\cite{BraandNor:2006}, to which the reader is referred for a fuller
account. The only differences here are the initial
magnetic field and the rotation, which are described in
sections~\ref{sec:rotation-setup} and \ref{sec:init-field}.
 
We use a three-dimensional MHD code developed by Nordlund \& Galsgaard (1995), which uses Cartesian coordinates, sixth-order spatial derivatives and the third-order predictor-corrector time-stepping routine of Hyman (1979), The code has a numerical diffusion scheme which is designed to damp structure on scales near the Nyquist frequency whilst preserving structure on well-resolved length scales. In the simulations presented here, a resolution of
$72^3$ was used. {\mk Tests were also performed at a lower resolution of $48^3$, and the results were not found to differ significantly, except at high azimuthal wavenumbers near the Nyquist frequency, as one would expect.}

{\mkb The fluid obeys the ideal gas equation of state $p=\rho RT$, and an energy equation is solved along with the mass and momentum conservation equations.
The star is modelled as a self-gravitating ball of gas, stably stratified (we use a polytropic of index n=3), embedded} in a
computational cube of side $4.5R_\ast$ where $R_\ast$ is the radius of
the star. Surrounding the star is a hot atmosphere of low electrical
conductivity {\mkb -- a physical magnetic diffusion term was added, in contrast to the low-level artificial scheme already present. This causes the field in the atmosphere to relax to a potential (curl-free) field over a timescale of the order of $R_\ast^2/\eta \sim 30\Omega^{-1}_{\rm crit}$ (where $\Omega_{\rm crit}$ is the break-up spin of the star, roughly equal to the reciprocal of the sound crossing time).} Boundary conditions are periodic.

\subsection{Rotation}
\label{sec:rotation-setup}

Instead of beginning the simulations with the star rotating inside the
computational box, we transfer to the rotating frame by adding a
Coriolis force. This avoids having shear flows at the boundaries,
which would be problematic in any geometry but particularly so in a
square computational box.

A strict transformation to the rotating
frame would of course also require a centrifugal force, but I have
not included this for the following three reasons. Firstly, it is not
thought that the centrifugal force affects the
stability of a magnetic field configuration, since, being a curl-free
force, it can be balanced by pressure forces, resulting merely in an
adjustment to the shape of the star, i.e. rotational
oblateness. Secondly, comparison with analytic results will be more direct,
as none of the earlier works included the centrifugal term. Thirdly,
a practical reason -- producing an initial equilibrium is easier with
a spherical star (although it can in principle be done in a flattened star). Eventually, one might want to add a centrifugal force to look at phenomena unique to stars rotating close to break up.

\subsection{Initial conditions}
\label{sec:init-field}

{\mk The non-magnetic aspect of the initial conditions is identical to that described in
detail in section 5 of \cite{BraandNor:2006}.} 
In the present study, we would like to create an equilibrium poloidal magnetic field to
start the simulation with. There are several configurations
to choose, and fortunately it is not expected that our choice will
affect the overall result (but this is confirmed later -- see
section~\ref{sec:other-fields}).

\begin{figure}
\includegraphics[width=1.0\hsize,angle=0]{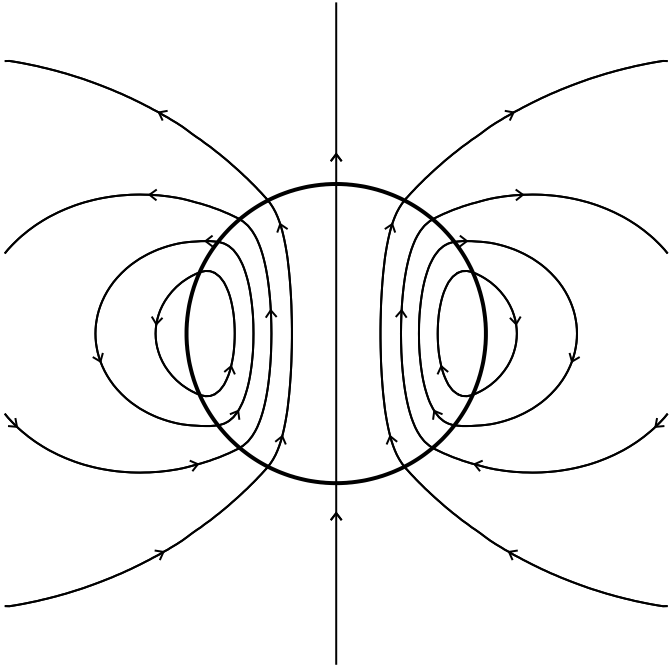}
\caption{The initial {\mkb equilibrium magnetic field, of the Roberts (1981) form.} The surface of the
star is represented by the thick line.}
\label{fig:gr-field}
\end{figure}

A first step towards finding an equilibrium field will be to find a
configuration where the Lorentz force
$\nabla\times\mathbf{B}\times\mathbf{B}$ is curl-free and can
therefore be balanced more easily by the pressure force. One example
of such a field is (\cite{Roberts:1981}):
\begin{eqnarray}
\mathbf{B}_r&=&\frac{B_0}{2}\cos\theta(5-3a^2), \nonumber\\
\mathbf{B}_\phi&=&0, \nonumber\\
\mathbf{B}_\theta&=&\frac{B_0}{2}\sin\theta(6a^2-5)\nonumber
\end{eqnarray}
inside the star, and
\begin{eqnarray}
\mathbf{B}_r&=&\frac{B_0\cos\theta}{a^3}, \nonumber\\
\mathbf{B}_\phi&=&0, \nonumber\\
\mathbf{B}_\theta&=&\frac{B_0\sin\theta}{2a^3} \nonumber
\end{eqnarray}
outside the star, expressed in spherical coordinates where
$a=r/R_\ast$, with $R_\ast$ being the stellar radius, and where $B_0$ is the
field strength at the surface of the star at the magnetic poles. We
shall use this configuration in the simulations. {\mk [This is the same field configuration as that used by Geppert \& Rheinhardt (2006).}] The field lines of
this configuration are plotted in fig.~\ref{fig:gr-field}. However,
simply adding this field to the star will not produce an
equilibrium. To reach a numerical equilibrium, the
simulation is run for a short time whilst the magnetic field is held fixed,
allowing the pressure {\mk (and density)} field{\mk(s)} in the star to adjust to the Lorentz force.

\subsection{Units and timescales}
\label{sec:timescales}

The maximum angular velocity at which a star can rotate without
breaking up is given by $\Omega_{\rm crit} =
\sqrt{GM_\ast/R_\ast^3}$ (where $M_\ast$ and $R_\ast$ are the stellar
mass and radius). The inverse of this, $\Omega_{\rm
crit}^{-1}$ -- which is roughly equal to the sound-crossing time -- I
use as the time unit in this paper. The mass and radius of the star are
used as the mass and length units, the results in
section~\ref{sec:results} being expressed in this system of units. In
these dimensionless units, the thermal energy of the star is equal to
$0.55$ and the magnetic energy used here is initially $1.3 \times 10^{-3}$.

The relevant frequencies (or inverse timescales) are
the angular velocity of the star $\Omega$ (which obviously cannot
exceed $1$ as this is the break-up velocity) and the Alfv\'en frequency $\omega_{\rm A}$, which is an average Alfv\'en speed
divided by the stellar radius and is given by $\omega_{\rm A} =
\sqrt{2E_{\rm mag}/M_\ast}/R_\ast$, where $E_{\rm mag}$ is the total
magnetic energy of the star. We are expecting the growth
rate of any instability present to be proportional and comparable to $\omega_{\rm A}$, {\mk or perhaps lower in the case of fast rotation (\cite{PitandTay:1985}).}

From observations of real stars we know that almost always
$\omega_{\rm A}\ll 1$, and that $\Omega$ may have any
value from $10^{-8}$ (e.g. some white dwarfs) to $0.9$ or higher
(e.g. Vega). In
the simulations presented here, $\omega_{\rm A} =
0.05$, which is high enough to be computationally practical whilst
still satisfying the condition $\omega_{\rm A}\ll 1$ and making both
the $\omega_{\rm A}\ll \Omega$ and $\omega_{\rm A}\gg \Omega$
regimes accessible.

\section{Results}
\label{sec:results}

First I shall describe simulations with parallel rotation and magnetic
axes ($\chi=0$), before coming to oblique rotators in
section~\ref{sec:oblique}.

\subsection{Aligned rotator with $\omega_{\rm A} = 0.05$}
\label{sec:aligned}

\begin{figure*}
\includegraphics[width=0.5\hsize,angle=0]{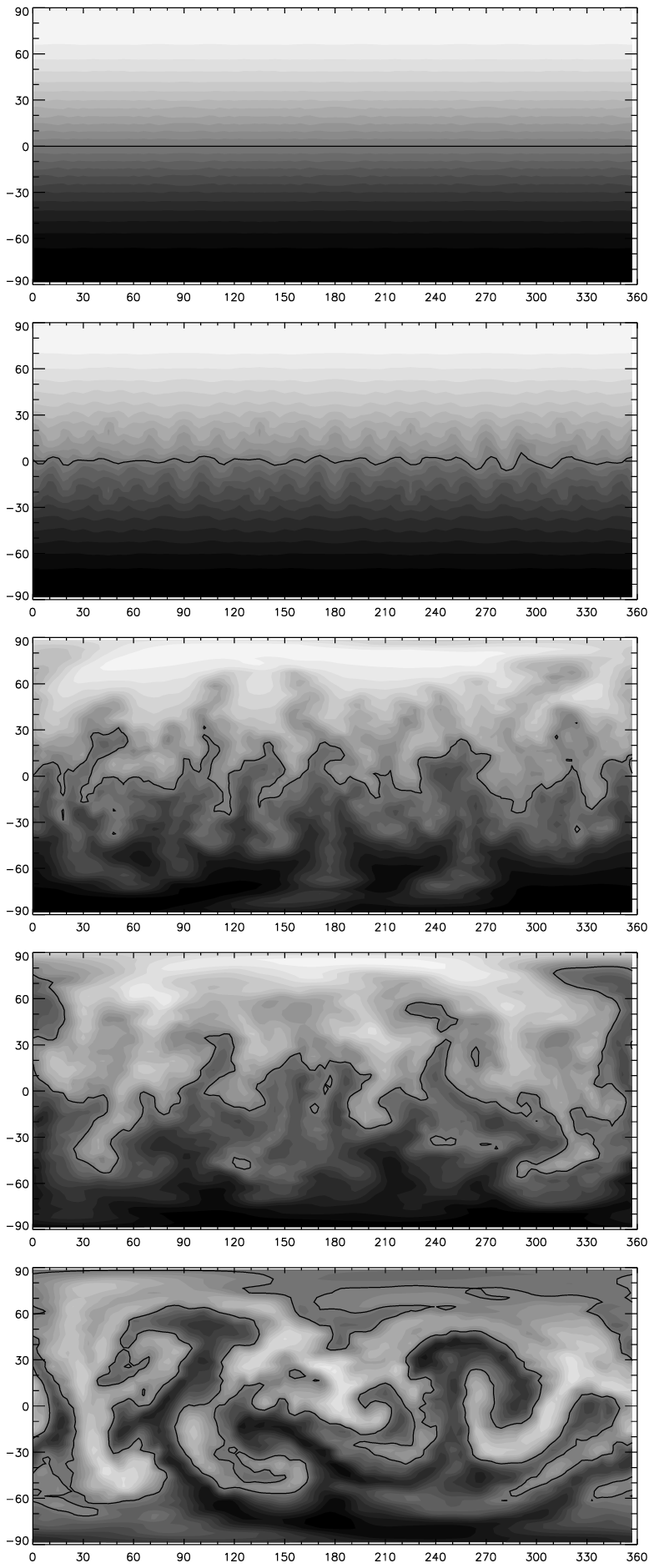}
\includegraphics[width=0.5\hsize,angle=0]{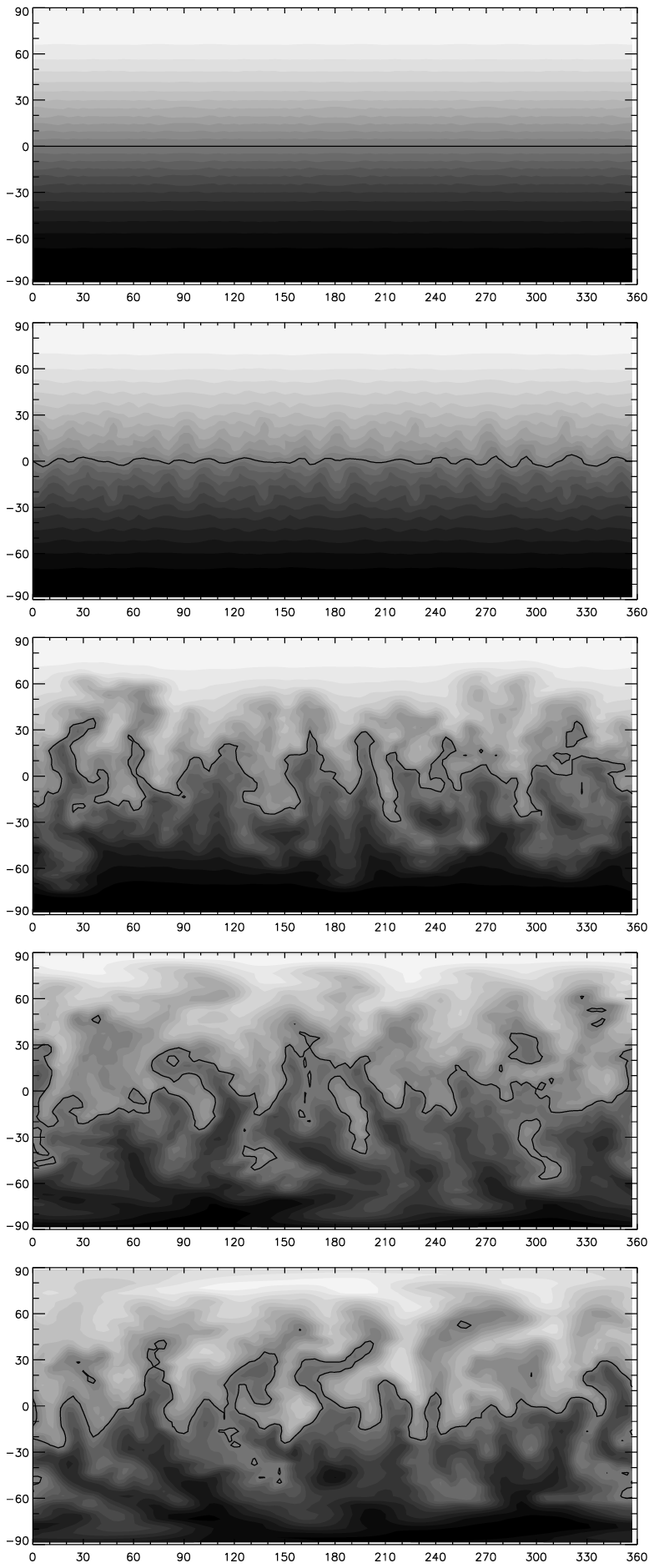}
\caption{Maps of the radial component of the magnetic field, $B_r$, on the surface of the star. White indicates positive $B_r$ and black negative; the $B_r=0$ line is also plotted. Longitude on the horizontal axis and Latitude on the vertical. Maps
are plotted from two runs, with $\Omega=0$ on the left and
$\Omega=0.5$ on the right, at five times $t=0, 40.1, 64.9, 92.8$ and
$154.7$. Rotation has no obvious effect on the linear development of
the instability, but at later times the rotation does restrict the
instability away from the equator.}
\label{fig:br-map}
\end{figure*}

To begin with, I run simulations with $\omega_{\rm A} = 0.05$ and a
variety of angular velocities $\Omega = 0, 0.125, 0.25, 0.5$ and
$1$. In fig.~\ref{fig:br-map} the radial component of the magnetic
field $B_r$ on the surface of the star is plotted at different
timesteps for two of the runs, $\Omega = 0$ and $0.5$. Looking first
at the left column of the figure, one can see the progress of the
instability in the non-rotating case: at first there is a movement of
gas both upwards and downwards, confined to the equatorial zone, and
the instability then spreads to the whole star, the field losing its
overall orientation. In the rotating case ($\Omega = 10 \omega_{\rm
A}$) we see that the instability is at first not affected, but that
instead of spreading unhindered to the whole star, it stays mainly in the equatorial
zone, the amplitude falling towards the poles. It seems then that rotation does not stabilise the magnetic
field, but merely restricts its growth.

\begin{figure}
\includegraphics[width=0.7\hsize,angle=0]{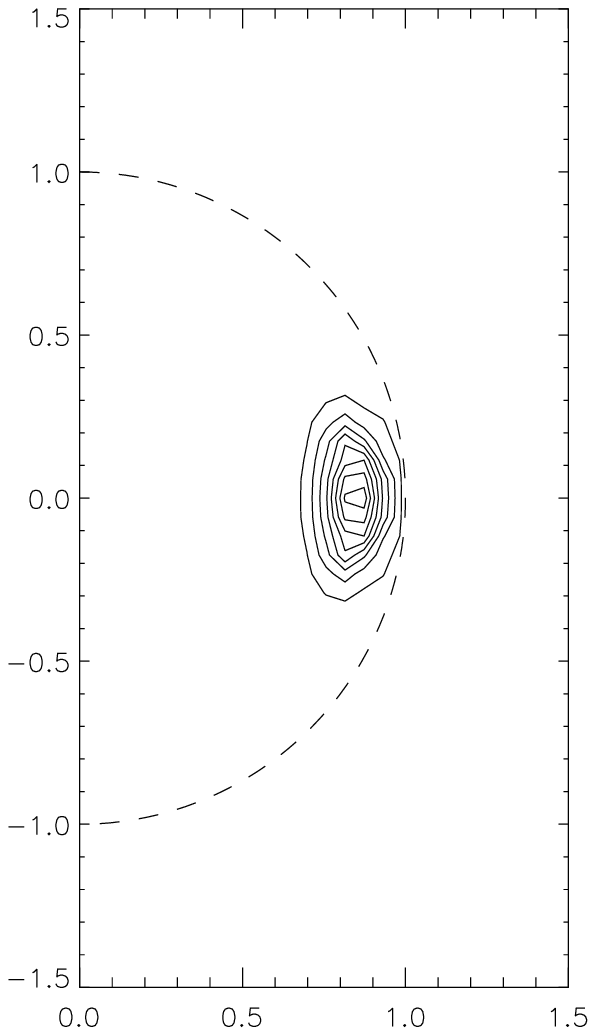}
\caption{Contour plot of the amplitude of the vertical component of the velocity field
$v_z$ in the $\varpi$-$z$ plane {\mkb (on the horizontal and vertical axes respectively)}, for the $m=13$ mode at time $t=30.3$
in the case where $\Omega = 0.5$. The dashed line
represents the surface of the star. It can be seen that the
instability is located on the equator of the star, just below the surface.}
\label{fig:vz-cross-section}
\end{figure}

We shall now look at the linear growth phase of the instability in
more detail. To do this, I first calculate the azimuthal Fourier components of the
vertical component of the velocity field $v_z$ (in cylindrical coordinates). {\mk The other components of the velocity field, or indeed of the magnetic field, can also be examined in this way, but $v_z$ is the most suitable because the amplitudes are higher.} The amplitude
of the $m=13$ mode in the $\varpi$-$z$ plane at a time $t=30.4$ in the
$\Omega=0.5$ case (before
the second plot on the right of fig.~\ref{fig:br-map}) is plotted in
fig.~\ref{fig:vz-cross-section}. This confirms our original picture of the
instability as being strongest on the magnetic equator.

\begin{figure*}
\includegraphics[width=0.5\hsize,angle=0]{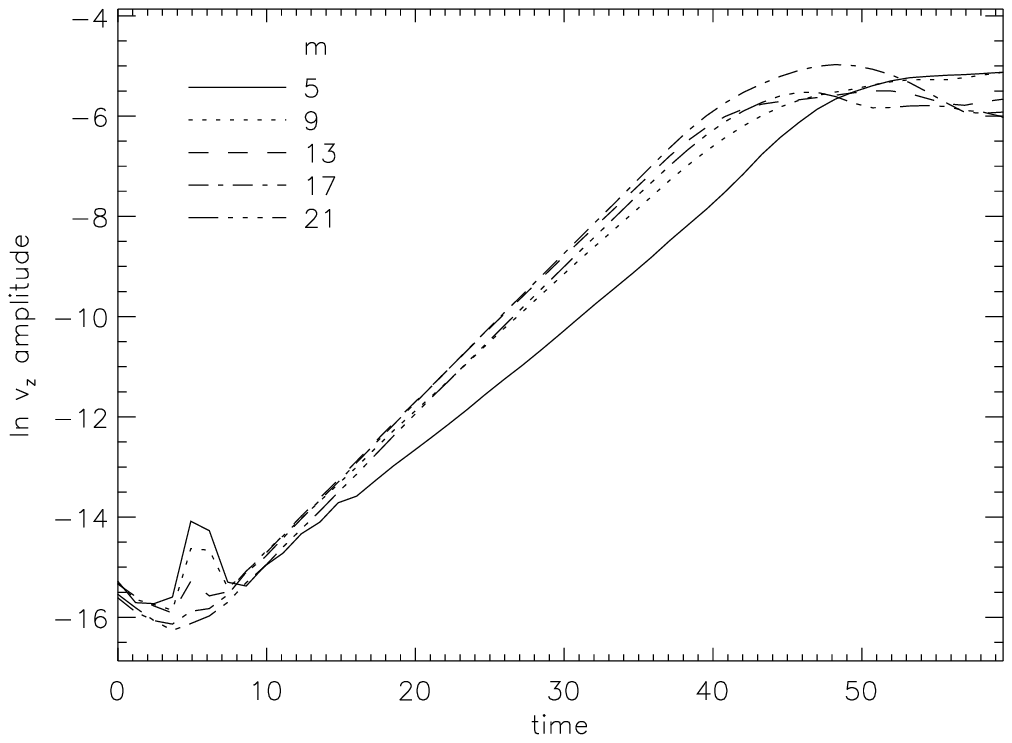}
\includegraphics[width=0.5\hsize,angle=0]{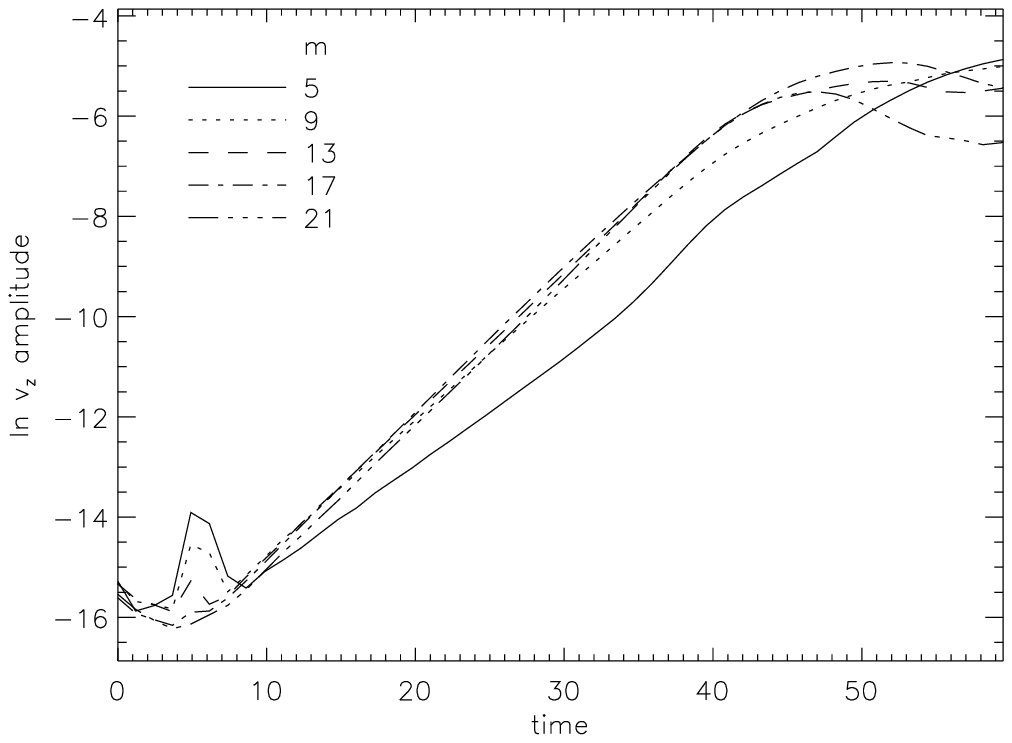}
\includegraphics[width=0.5\hsize,angle=0]{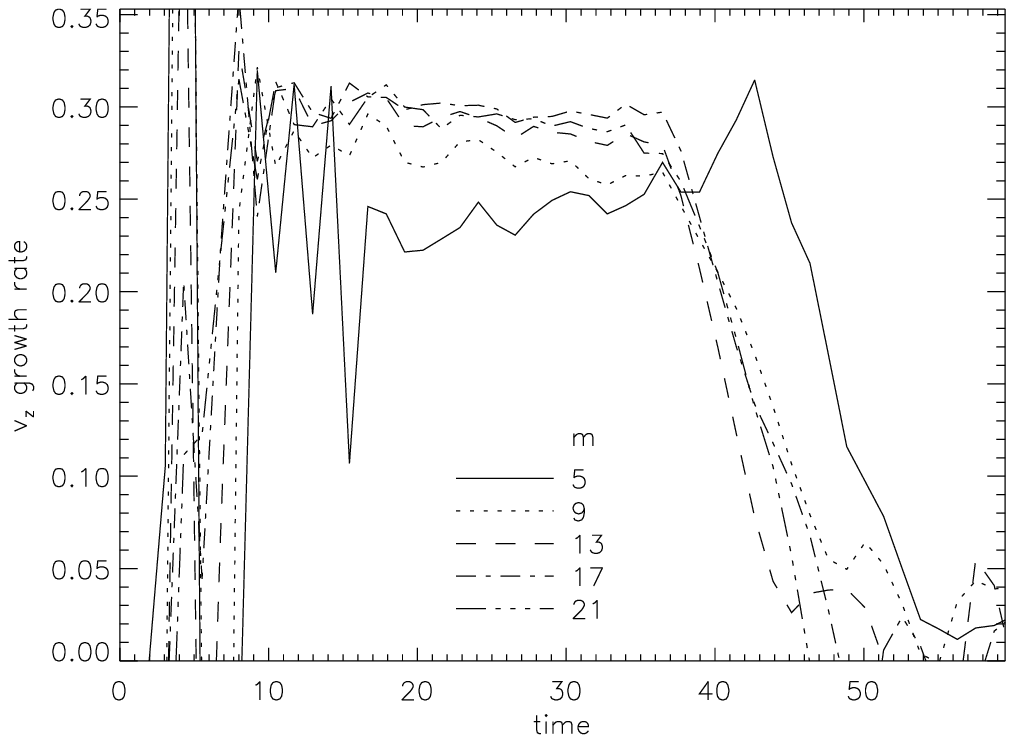}
\includegraphics[width=0.5\hsize,angle=0]{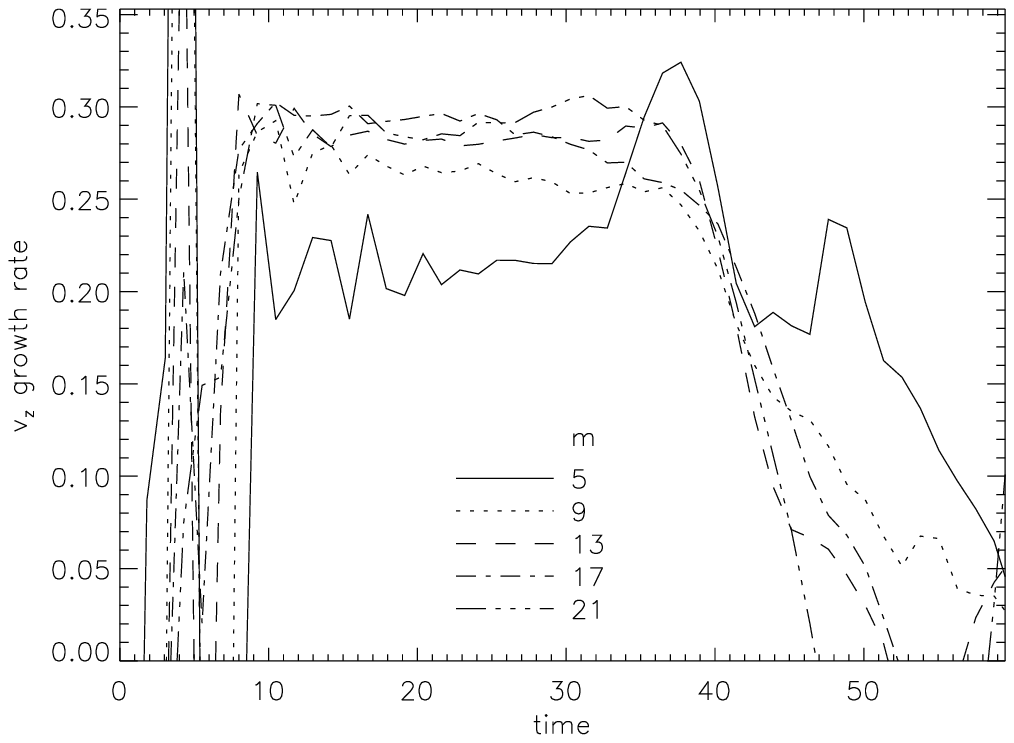}
\caption{Above, amplitudes of five azimuthal modes for the non-rotating case
(above left) and for the $\Omega=0.5$ case (above right). {\mk Aligned ($\chi=0$) case, Roberts initial field.} These
plots end at the time when the instability growth becomes
non-linear. Below, the time derivatives of these {\mk log} amplitudes, i.e. the
growth rates of the various $m$ modes. It is clear that rotation has little effect on the linear stage.}
\label{fig:lnvz-vzgr}
\end{figure*}

We can now integrate
these azimuthal modes over an area in the $\varpi$-$z$ plane (from $\varpi=0.6$ to $0.9$
and from $z=-0.15$ to $+0.15$, i.e. where the velocity field is
strongest), and the resulting amplitudes of five Fourier modes are
plotted in fig.~\ref{fig:lnvz-vzgr}, along with their associated time
derivatives, for the $\Omega=0$ and $0.5$ cases. It can be seen that
after some initial fluctuations, the modes grow
fairly steadily. All of the modes grow at roughly the same growth rate,
around $0.28$ or $5.6\omega_{\rm A}$, with just the $m=5$ mode lagging
behind; a little higher than we were expecting, but it is
safe to assume that this is simply to do with our definition of
$\omega_{\rm A}$: in the part of the star where the instability is
strongest, the Alfv\'en velocity is higher than the average for the
whole star. {\mkb This is because the density is much lower there (by a factor $\sim 1000$) than in the centre of the star, more than making up for the somewhat weaker magnetic field.} More importantly, there is very little difference between the non-rotating and
rotating cases in this linear growth phase.

\begin{figure}
\includegraphics[width=1.0\hsize,angle=0]{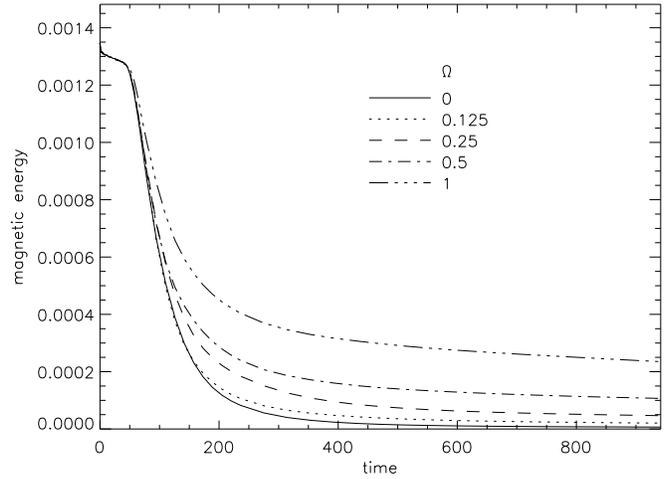}
\caption{Total magnetic energy against time for runs with different
values of $\Omega$: $0, 0.125, 0.25, 0.5$ and $1$. It can be seen that
rotation merely slows the decay rather than stopping it altogether. {\mk Aligned ($\chi=0$) case, Roberts initial field. }}
\label{fig:me-o}
\end{figure}

\begin{figure}
\includegraphics[width=1.0\hsize,angle=0]{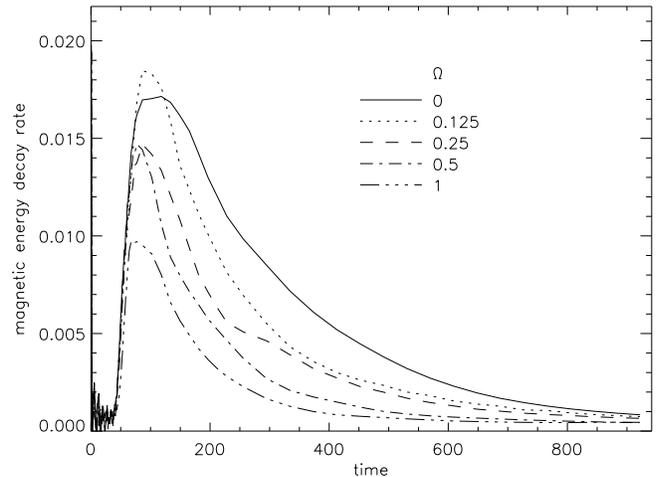}
\caption{Rate of decay of magnetic energy against time for $\Omega =
0, 0.125, 0.25, 0.5$ and $1$, i.e. (minus) the time derivative of (log) fig.~\ref{fig:me-o}.}
\label{fig:me-decay-o}
\end{figure}

Rotation therefore only affects the non-linear development of the
instability, and it is this which is most interesting for applications
in astrophysics because it is during the non-linear phase that the
magnetic energy is lost. In fig.~\ref{fig:me-o} the total magnetic
energy in the star is plotted against time, for the five runs with
$\Omega = 0, 0.125, 0.25, 0.5$ and $1$. One can see that in all cases,
most of the energy is destroyed by the instability. However, it is not
clear from this graph alone that the total magnetic energy is
heading to zero. The rate at which magnetic energy is being lost,
$-(1/E_{\rm mag})dE_{\rm mag}/dt$ is plotted in
fig.~\ref{fig:me-decay-o}, showing that the rate of energy loss
reaches a peak and then declines. However as the instability, and
therefore the rate of energy loss, operate at a speed dependent on the
Alfv\'en frequency, we should of course expect the magnetic decay rate to decline.

\begin{figure}
\includegraphics[width=1.0\hsize,angle=0]{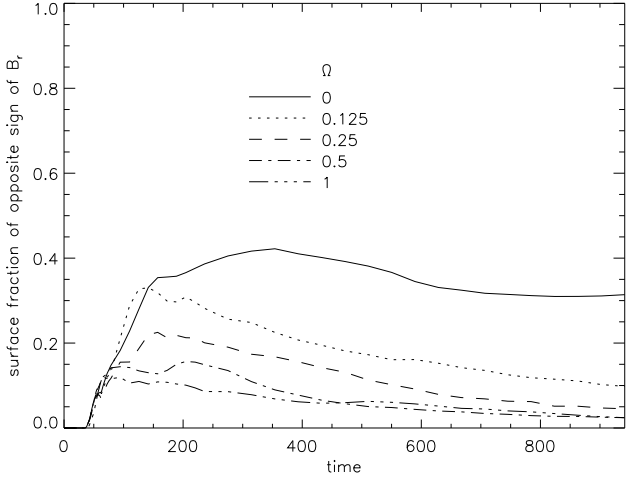}
\caption{The fraction of the surface of the star where the radial
component of the magnetic field $B_r$ is of opposite sign to
that initially, for five runs with different rotation speeds. In the
non-rotating case, this quantity stays close to $0.5$, but in all of
the other cases, it sinks slowly back down to zero after the initial
rise. This is because the instability is largely confined to a volume near the
equatorial plane, and as the magnetic energy is destroyed there,
$\omega_{\rm A}$ falls, and the
thickness of this volume falls too.}
\label{fig:mapfrac-o}
\end{figure}

It is possible to quantify the containment of the instability to the
equatorial region by defining a new quantity as the fraction of the
stellar surface over which the sign of $B_{\rm r}$ is not the same as
at $t=0$. Evolution to a completely random field distribution would give a value around
$0.5$, while a configuration like that in the lower-right of
fig.~\ref{fig:br-map} would have a smaller
value. Fig.~\ref{fig:mapfrac-o} is a graph of this quantity against
time for the five runs with different angular velocities. It can be
seen that in the four rotating runs, the value sinks gradually back
down to zero. This is because the Alfv\'en frequency $\omega_{\rm A}$
falls, also falling in relation to the angular velocity
$\Omega$. This in turn causes the more unstable region to shrink towards
the magnetic equator, so there is the double effect of energy being
lost both more slowly and in a smaller volume. {\mkb Eventually, we would expect the quantity plotted in fig.~\ref{fig:br-map} to fall to zero as the total magnetic energy falls to zero. Also, note that in numerical simulations, in practice the rate of energy loss from the instability will eventually fall below the energy loss rate from Ohmic dissipation of the global magnetic field, and the instability will no longer have much effect on the field's evolution. In a real star, with much lower magnetic diffusivity than is accessible in simulations, this end-state will be reached only when the field has become vanishingly weak.}

\begin{figure}
\includegraphics[width=0.32\hsize,angle=0]{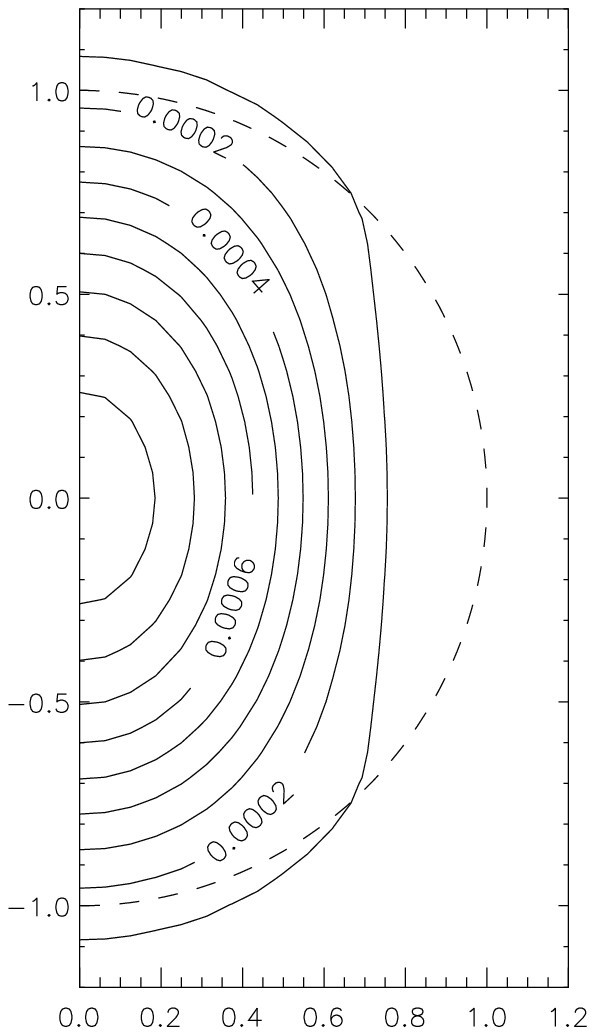}
\includegraphics[width=0.32\hsize,angle=0]{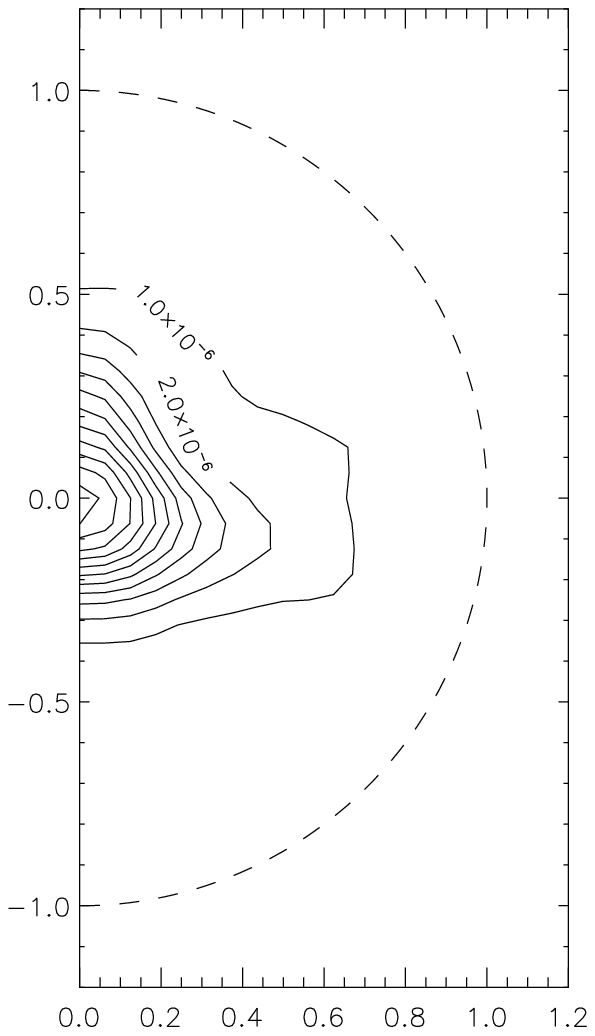}
\includegraphics[width=0.32\hsize,angle=0]{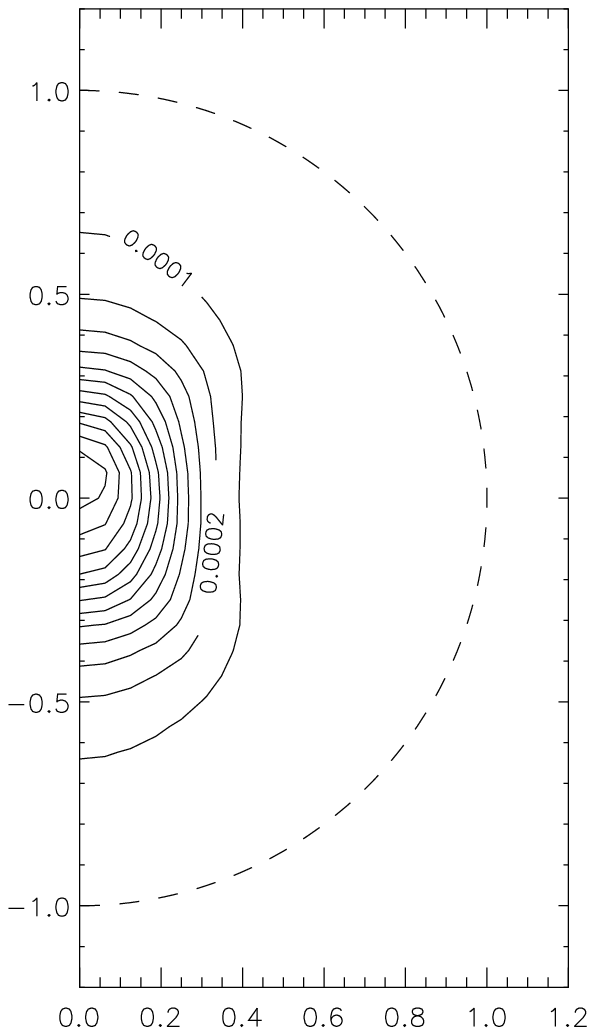}
\caption{An azimuthal average {\mk in the $\varpi-z$ plane} of the magnetic energy density
$B^2/8\pi$, at time $t=0$ (left), at time $t=926$ in the $\Omega=0$
case (centre), and at time $t=926$ in the $\Omega=1$
case (right). The dotted line represents the stellar surface. {\mk Aligned ($\chi=0$) case, Roberts initial field.}}
\label{fig:xsec-hb2}
\end{figure}

\begin{figure}
\includegraphics[width=0.32\hsize,angle=0]{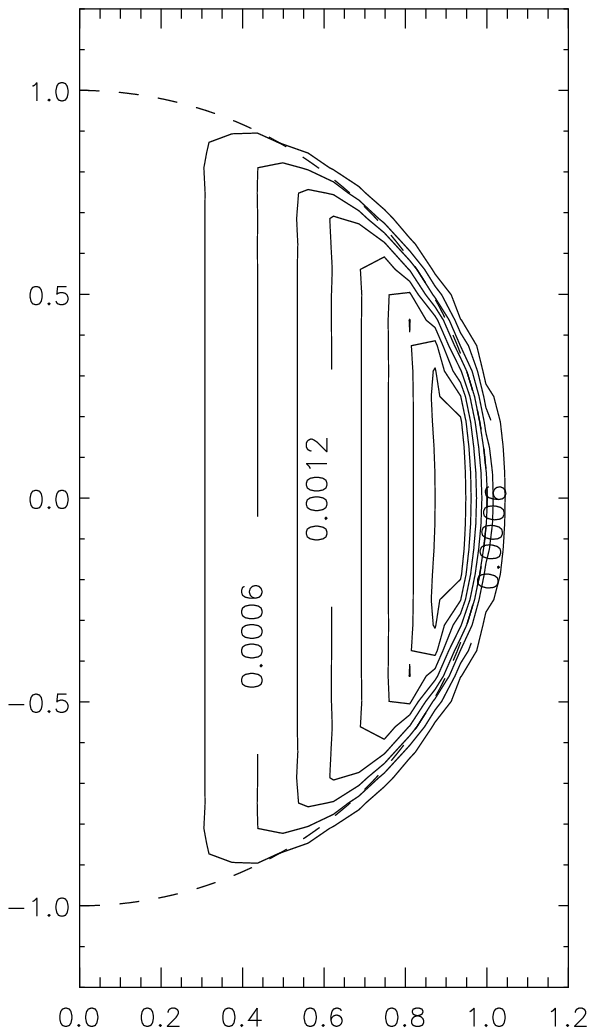}
\includegraphics[width=0.32\hsize,angle=0]{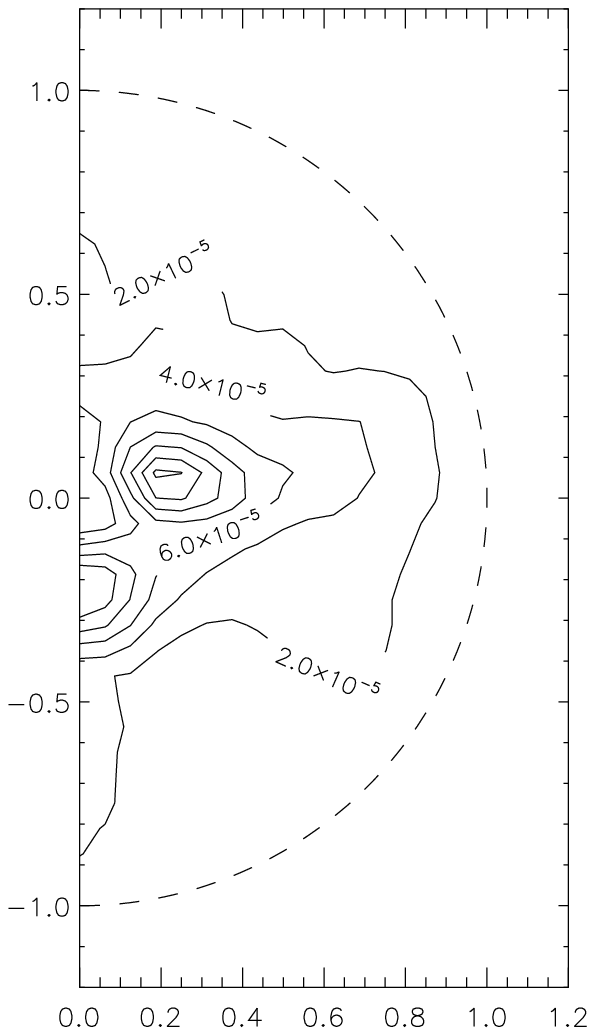}
\includegraphics[width=0.32\hsize,angle=0]{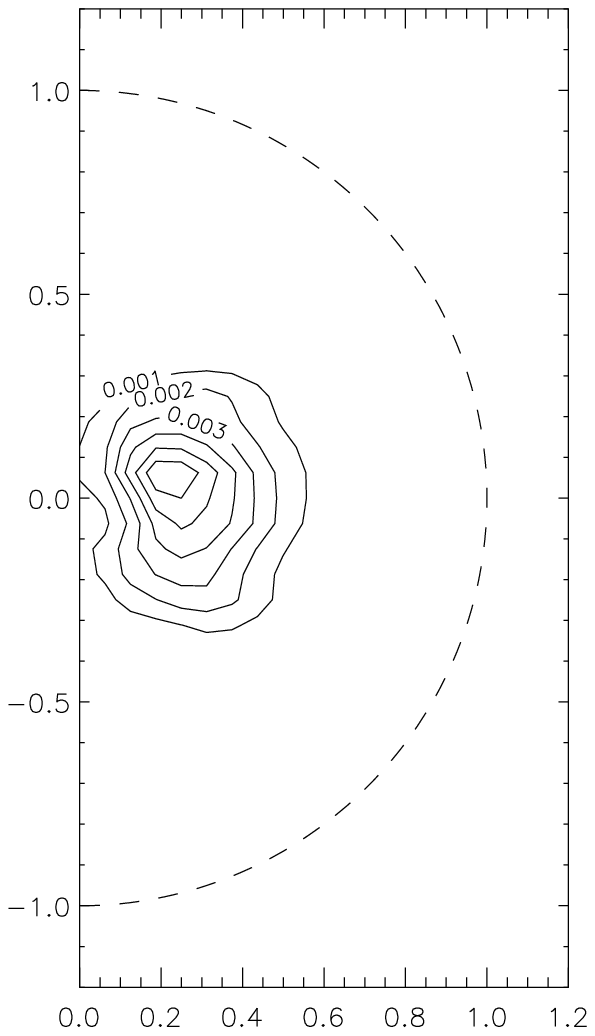}
\caption{An azimuthal average {\mk in the $\varpi-z$ plane} of the current density
$J^2$, at time $t=0$ (left), at time $t=926$ in the $\Omega=0$
case (centre), and at time $t=926$ in the $\Omega=1$
case (right). The dotted line represents the stellar surface. {\mk Aligned ($\chi=0$) case, Roberts initial field.}}
\label{fig:xsec-j2}
\end{figure}

\begin{figure}
\includegraphics[width=1.0\hsize,angle=0]{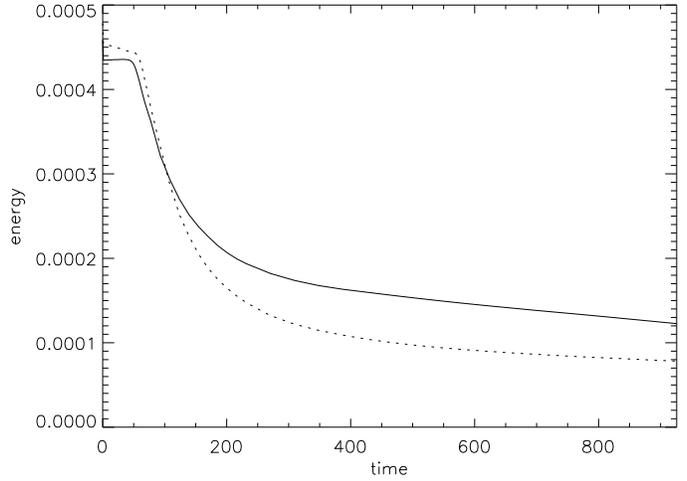}
\caption{Magnetic energy inside the star, in two components: the
volume within $0.3R_\ast$ (solid line) of the equator {\mk (i.e. $-0.3R_\ast<z<0.3R_\ast$)} and the rest of
the star (dotted line). {\mk Aligned ($\chi=0$) case, Roberts initial field, $\Omega=1$.}}
\label{fig:energy-mid}
\end{figure}

\begin{figure}
\includegraphics[width=1.0\hsize,angle=0]{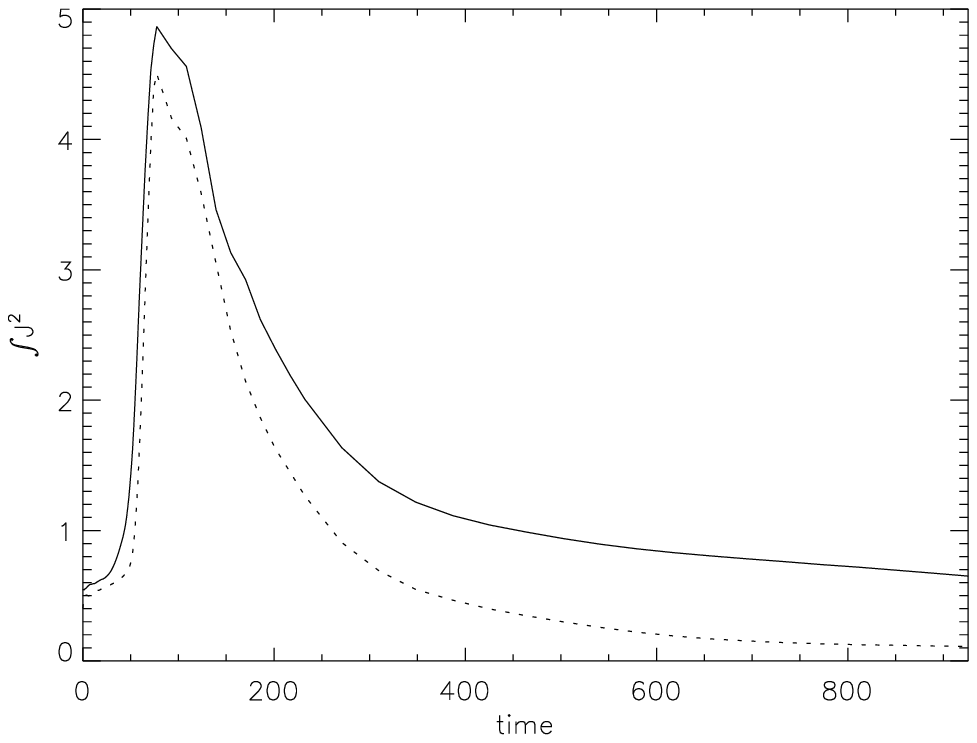}
\caption{Integrated current density inside the star, in two components: the
volume within $0.3R_\ast$ (solid line) of the equator {\mk (i.e. $-0.3R_\ast<z<0.3R_\ast$)} and the rest of
the star (dotted line). {\mk Aligned ($\chi=0$) case, Roberts initial field, $\Omega=1$.}}
\label{fig:joule-mid}
\end{figure}

Figs.~\ref{fig:xsec-hb2} and \ref{fig:xsec-j2} show azimuthal
averages of the magnetic energy density and current density at
different times in the non-rotating and rotating cases. It can be seen
that the region of strongest field shrinks in both cases towards the
centre of the star, and that the region where the magnetic energy is
being lost (where the current density is high) is restricted in the
rotating case to a ring around the equator. It is also possible to
look at the energy balance between the equatorial region and the rest
of the star. For the rotating case $\Omega=1$,
figs.~\ref{fig:energy-mid} and \ref{fig:joule-mid} show
magnetic energy and integrated current density plotted against time,
divided into two parts: the equatorial zone (defined as
$-0.3R_\ast<z<0.3R_\ast$) and the rest of the star. These graphs show
that although the equatorial zone accounts for the majority of
magnetic energy loss, the energy outside of this zone decays just as
fast or faster as inside the equatorial zone. From this we may infer
the presence of some kind of transport process carrying energy from
the polar regions towards the equator.

\subsection{The oblique rotator}
\label{sec:oblique}

\begin{figure}
\includegraphics[width=1.0\hsize,angle=0]{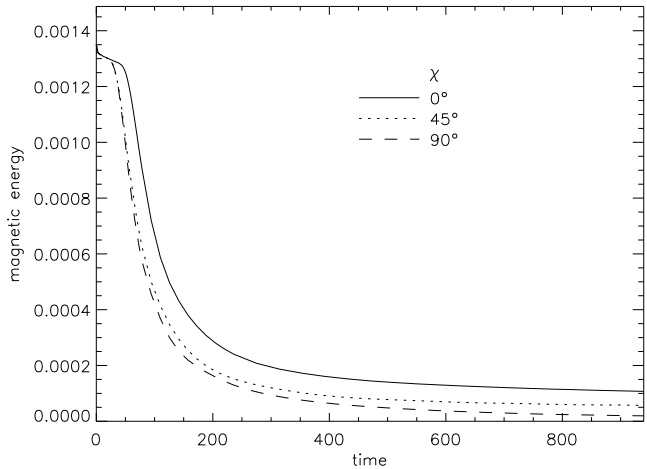}
\caption{Total magnetic energy against time for runs with different
values of $\chi$: $0^\circ, 45^\circ$ and $90^\circ$. The magnetic
field is unstable, regardless of the relative alignment of the
magnetic and rotation axes. {\mk Roberts initial field, with $\Omega=0.5$.}}
\label{fig:me-chi}
\end{figure}

If $\chi$ is the angle between the rotation axis and magnetic axis, we
have seen that if $\chi=0$ the rotation is not able to stabilise the magnetic field. We shall now
generalise to $\chi \not= 0$. Identical simulations were run to the
$\Omega=0.5$ run described in section~\ref{sec:aligned}, but with
$\chi=45^\circ$ and $90^\circ$. The magnetic energy in these runs
(along with the original $\chi=0^\circ$ run) is plotted in
fig.~\ref{fig:me-chi}. It is clear that the field is still not
stabilised -- in fact, the magnetic energy falls slightly more quickly than when
the two axes are aligned.

\subsection{Other poloidal field configurations}
\label{sec:other-fields}

\begin{figure}
\includegraphics[width=1.0\hsize,angle=0]{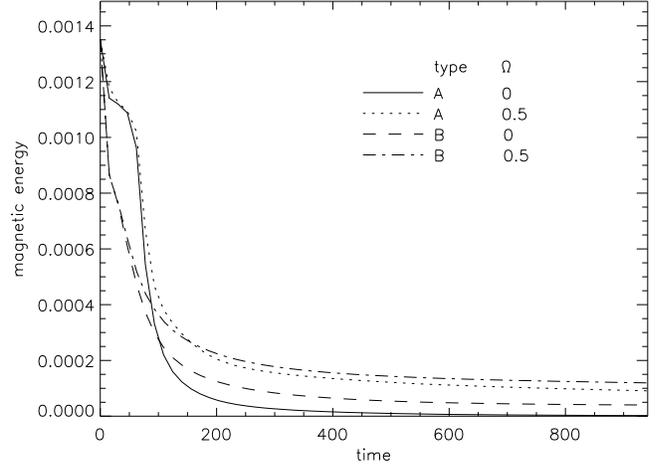}
\caption{Total magnetic energy of the star against time. Four
runs, two with each kind of magnetic field configuration, of which one
is non-rotating and one rotating at $\Omega=0.5$. {\mk Aligned ($\chi=0$) case.} The magnetic energy
falls right at the beginning as a numerical equilibrium is found,
before the instability sets in to destroy more magnetic energy. }
\label{fig:me-other}
\end{figure}

\begin{figure}
\includegraphics[width=1.0\hsize,angle=0]{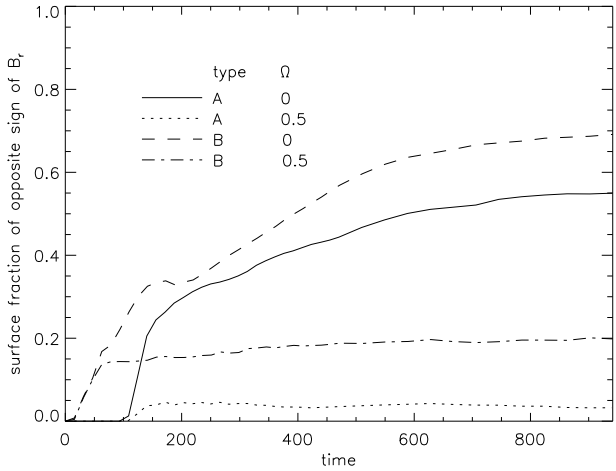}
\caption{The fraction of the surface of the star where the radial
component of the magnetic field $B_r$ is of opposite sign to
that initially, for the four runs in fig.~\ref{fig:me-other}. It can
be seen that the instability grows faster in case B. {\mk Aligned ($\chi=0$) case.}}
\label{fig:mapfrac-other}
\end{figure}

It is now necessary to repeat these simulations with some different
poloidal field configurations, in case the field used so far is
special in some way. We shall use two alternative poloidal
fields. First, a field (the poloidal component of a field suggested by
Kamchatnov, 1982; denoted as type A in figs.~\ref{fig:me-other} and \ref{fig:mapfrac-other}) given by:
\begin{eqnarray}
\mathbf{B}_r&=&\frac{B_0 \cos{\theta}}{(1+a^2)^2}, \nonumber\\
\mathbf{B}_\phi&=&0, \nonumber\\
\mathbf{B}_\theta&=&\frac{B_0 \sin{\theta} (a^2-1)}{(1+a^2)^3}\nonumber
\end{eqnarray}
where $a=r/R_{\rm m}$, with $R_{\rm m}$ being some radius less than
the radius of the star. In the following simulations, I use $R_{\rm
m}=0.25R_\ast$. Secondly, I use a magnetic field uniform in strength
and direction inside the star attached to a curl-free field outside
the star, as on the right-hand-side of fig.~\ref{fig:stable-and-ums}
(denoted as type B in figs.~\ref{fig:me-other} and
\ref{fig:mapfrac-other}). This is the field configuration used to
study the non-rotating case by Braithwaite \& Spruit (2006).
{\mk These two configurations were picked for the following reasons. The first of these fields has the property that most of the field lines close inside the star, in contrast to the field used up to now, where most of the field lines go through the surface of the star. Moreover, it is found that together with the toroidal component, as proposed by Kamchatnov, it can be stable (in conjunction with the right velocity field) and has basically the same form as the stable fields found in \cite{BraandNor:2006}. Indeed, simulations using the Kamchatnov field (with both poloidal and toroidal components) as the initial field do appear to show that the field is stable in a non-rotating star, or at least that the field quickly relaxes into a very similar-looking stable state. [This will be described in more detail in a forthcoming paper.] Note that the field as given by the equations above does contain a non-potential field in the atmosphere, which will quickly relax to curl-free at the beginning of the simulation. The second field was chosen because of its simple structure, because all field lines
close outside the star, and because it has already been tested in non-rotating stars in \cite{BraandSpr:2006}.}

The simulations described in section~\ref{sec:aligned} were run again
with these field configurations. The same instability was found,
although the growth rates differed by some factors of order
unity. Again, rotation failed to stabilise the field, merely
restricting the non-linear development of the instability. The
magnetic energy in some of these runs is plotted in
fig.~\ref{fig:me-other}, and the fraction of the stellar surface over
which the sign of $B_r$ has changed is plotted in
fig.~\ref{fig:mapfrac-other}. Although the magnetic energy, and
therefore $\omega_{\rm A}$ as defined in section
\ref{sec:timescales}, are the same (and the same as in previous runs), the instability grows faster in case B,
probably because it is located near the stellar surface where the
Alfv\'en speed is higher than nearer the centre of the star in case
A. Owing to this higher growth rate, rotation has less
effect on case B, or rather, a higher rotation speed is required to
slow the field decay to the same extent as in case A, since it is the
ratio of growth rate of the instability to the star's rotation rate
that determines the effect of the rotation.

\section{Conclusions and discussion}
\label{sec:concs}

It had previously been established that a purely poloidal magnetic
field in a non-rotating star is subject to a magnetohydrodynamic
instability whose growth timescale is comparable to the Alfv\'en
crossing time (Markey \& Tayler 1973, 1974, \cite{Wright:1973},
\cite{BraandSpr:2006}). In this paper, I have examined the case of
a poloidal magnetic field in a rotating star, by following the
evolution of the magnetic field numerically. All three different poloidal
configurations tried were unstable and did not reach a stable state,
even when $\Omega=20\omega_{\rm A}$. There is no obvious reason to
suppose that stabilisation is provided at even higher rotation speeds,
since the rotation up to that level is shown to have no effect on the
linear growth phase of the instability. Nor is there any reason to
believe that some other kind of poloidal field from the three tried
here would be stable. In addition, no stable state
was found by changing the angle between the magnetic and rotation
axes.

{\mk The result that the rotation has no effect at all on the linear
growth rate of the instability is different to the corresponding
result in toroidal fields, where rotation {\mkb is generally expected to} reduce the
growth rate by a factor of order $\omega_{\rm A}/\Omega$ {\mkb (\cite{PitandTay:1985})}. This can be
explained by thinking in terms of stable stratification, which  prevents any significant displacements in the direction
of the gravitational field, and the
Coriolis force $-2\mathbf{\Omega}\times\mathbf{v}$. In poloidal fields aligned with the
rotation axis, the instability is located on the equator, and gravity
allows displacements in the $z$ direction (in cylindrical polar
coordinates), which are not affected by the Coriolis term, and in the
$\phi$ direction, which produce a Coriolis force in the $\varpi$
direction which {\mkb can then be largely} balanced by gravity. In the oblique rotator,
there is still at least part of the (magnetic) equator where the Coriolis term
has no effect. With a toroidal field where the magnetic and rotation axes
are aligned, on the other hand, gravity and rotation are parallel {\mkb on the magnetic axes}, so
the Coriolis force acts on all displacements allowed by the
stratification {\mkb (except those exactly on the equator)}. However, in the case where the toroidal magnetic axis
and rotation axis are perpendicular, the rotation again has no effect. This is analogous to the absence of Coriolis effects in the equatorial part of the Earth's atmosphere, where for example no high-pressure anticyclones are possible, only centrifugally-supported tropical typhoons.}

{\mk The results should be of relevance to a variety of magnetic
stars. In the case of Ap stars, we see a range of magnetic field
strengths $300$ G to $30$ kG and periods $12$ hours to $100$
years. Calculating the ratio $\Omega/\omega_{\rm A}$ we find values of
$0.3$ to $10^5$, assuming that the field strength on the surface
is representative of that in the interior; if the field is stronger in
the interior than on the surface, these values of $\Omega/\omega_{\rm
A}$ would be lower. So some of the Ap stars (those with a ratio below
1) could never be stabilised by rotation anyway, but in the majority
the rotation would be fast enough for stabilisation, if it were
possible in principle. Magnetic white dwarfs have fields of $10^5$ to $10^9$ G,
periods of a few minutes to $100$ years and $\Omega/\omega_{\rm A}$
ratios of $3\times 10^{-4}$ to $10^5$, so are even more clearly split
between the fast and slow rotating regimes. So in both of these classes
of object, some of the stars must have a stable field without
rotational stabilisation. Also, since no fundamental difference in
magnetic topology has been observed between the quickly and slowly rotating
cases (see, e.g. \cite{LanandMat:2000}), there is no evidence that rotation
allows the existence of different field configurations. {\mkb [However, since it is generally not possible to observe a toroidal field component, it cannot be ruled out that the field configurations of slow and fast rotating Ap stars and magnetic WDs do differ in their toroidal parts, although it seems unlikely that the poloidal components would show no difference in this case.]} Of the
neutron stars, the Anomalous X-ray Pulsars and Soft Gamma Repeaters
(the ``magnetar candidates'') with spindown-inferred fields of
$10^{14}$ to $10^{15}$ gauss, are in the slowly rotating regime {\mkb (at least at the ages at which we observe them -- opinions differ as to possible spin periods at birth)}, while
the radio pulsars are in the fast rotating regime. Neutron stars have
the added complication of a solid crust which could in principle
stabilise an otherwise unstable field. The strength of this crust is
somewhat uncertain but is estimated to be strong enough to stabilise
pulsar fields but not magnetar fields, although in pulsars one still
has to deal with the issue of the time taken for the crust to
freeze. In all examples of all three types of star, if the field were
unstable and decayed on a timescale of $1/\omega_{\rm A}$ or even
$\Omega/ \omega_{\rm A}^2$, the field would decay on a timescale very
short compared to the star's lifetime, and we would not observe any magnetic
field.}

On the basis of these results, that there is no evidence for the
existence of a stable poloidal field in a rotating star, and because
we already know that purely toroidal fields are also unstable, we must
look for fields of a mixed poloidal-toroidal geometry to explain
observations of magnetic fields in stars not thought able to
accommodate any regenerative dynamo process, even those stars where
the rotational timescale $\Omega^{-1}$ is much shorter than the
Alfv\'en crossing time.

Purely toroidal and purely poloidal fields have zero magnetic
helicity. The mixed toroidal-poloidal torus field found in
\cite{BraandSpr:2004} and \cite{BraandNor:2006} has non-zero helicity
(in a sense helicity is the product of the toroidal and poloidal
components) and is dynamically stable. It seems likely therefore that
there is no simple configuration with zero helicity which is
stable. If a stable zero-helicity configuration does exist, it is
probably a composite of two or more parts, each with non-zero
helicity.  In this study it was only possible to follow the evolution
of the magnetic field for a certain length of time, making it
impossible to check the following mechanism. It is conceivable that
during the decay of the field, some helicity leaks through the surface
of the star, in a random manner. Despite contributions from different
parts of the stellar surface largely cancelling each other out,
especially when the dominant wavenumbers are high, the field inside
the star could acquire a net magnetic helicity. This would enable the
field to find its way into a stable configuration of mixed
poloidal-toroidal form, instead of decaying indefinitely. However, the
strength of this residual field would be extremely low compared to its
initial strength.

\begin{figure}
\includegraphics[width=1.0\hsize,angle=0]{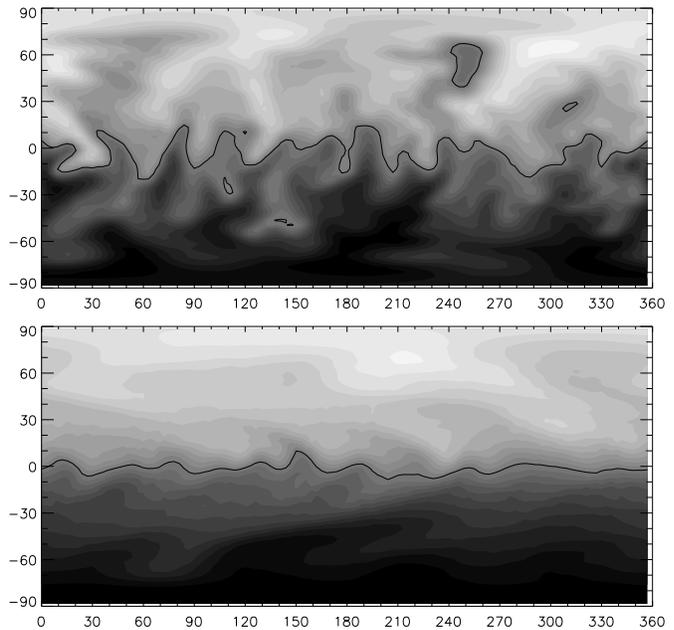}
\caption{Maps of the radial component of the magnetic field, $B_r$, on
the surface of the star (as fig.~\ref{fig:br-map}), in the
aligned-rotator case, Roberts field, with $\Omega=0.5$, at two times:
$t=386.9$ and $925.1$. Clearly, the amplitude of the instability away
from the equator decreases as the field energy and therefore Alfv\'en frequency fall.}
\label{fig:latestage}
\end{figure}

\begin{figure}
\includegraphics[width=1.0\hsize,angle=0]{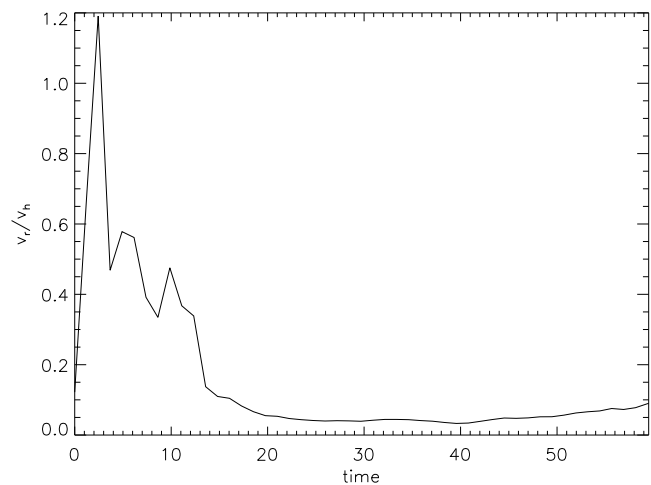}
\caption{{\mkb The fraction $|v_r|/\sqrt{v_\phi^2+v_\theta^2}$ (of the $m=13$ mode, integrated over the interior of the star in the $\theta, \phi$-plane) against time. It is clear that, after a short early phase, the radial (parallel to gravity) displacements are indeed much less than the horizontal (perpendicular to gravity) displacements.}}
\label{fig:vr_frac}
\end{figure}

This study finds that rotation has no effect on the linear growth of
the instability in poloidal fields, {\mkb and that its effect is
limited to slowing (but not stopping entirely) the decay of the field in the non-linear phase -- the amplitude away from
the equator is much reduced.} This result does not agree with
those of Geppert \& Rheinhardt (2006), who find that rotation can
stabilise the field entirely. The origin of this contradiction is not
certain. However, some possibilities spring to mind.  First, they use
different physics -- an incompressible {\mkb fluid}, as opposed to a
compressible fluid in this study. {\mk There is therefore no stable
stratification, and one would normally expect the field to be {\it
less} stable. {\mkb Fig.~\ref{fig:vr_frac} shows that the displacements are very nearly perpendicular to gravity, as one would expect in a stably stratified star -- so it looks like the compressibility is not important here anyway.}
Another possible reason for the discrepancy is as follows.
In the rotating case the instability
has greatest amplitude in the equatorial zone, and as the field decays
and the ratio $\Omega/\omega_{\rm A}$ increases {\mkb this effect, of
greater amplitude nearer the equator than further from the equator, also}
increases.
As this happens, the amplitudes of the azimuthal $m>0$ modes
decrease, and it is difficult to distinguish decay via instability
from decay via Ohmic diffusion, due to insufficient separation of the
two timescales in the simulation. This is illustrated in
fig.~\ref{fig:latestage}, where $B_r$ on the stellar surface is
plotted a long time after the instability has become unstable and most
of the  magnetic energy has been lost. Note also that if the ratio
$\Omega/\omega_{\rm A}$ is very high from the beginning, it may not be
possible to see very much happening if the diffusivity is too high. In
a numerical simulation the diffusivity is often too high, as it is
inevitably several orders of magnitude higher than in a real star, and
so we may expect that in nature, the magnetic field of a star would
decay to an extremely low level (on a timescale short compared to a
stellar evolution timescale) before the decay rate from instability
falls to that from global Ohmic diffusion.}

{\it Acknowledgements.} The author would like to thank Henk Spruit for
valuable discussions and suggestions, {\mk as well the the referee, Matthias Rheinhardt, for comments which led to significant improvements in the paper.}

\begin{appendix}

\end{appendix}



\begin{thebibliography}{99}

\bibitem[Angel et al. 1974]{Angeletal:1974}Angel, J.R.P., Carswell, R., Strittmatter, P.A., Beaver, E.A. and
Harms, R. 1974, ApJ, 194, L47.
\bibitem[Babcock 1947]{Babcock:1947}Babcock, H.W., 1947,
Astrophys. J., 105, 105.
\bibitem[Borra \& Landstreet 1980]{BorandLan:1980}Borra, E.F. and Landstreet, J.D., 1980, ApJSuppl, 42, 421.
\bibitem[Braithwaite 2006]{Braithwaite:2006a}Braithwaite,
J., 2006, A\&A, 453, 687.
\bibitem[Braithwaite \& Nordlund 2006]{BraandNor:2006}Braithwaite, J.
and Nordlund, \AA, 2006, A\&A, 450, 1077.
\bibitem[Braithwaite \& Spruit 2004]{BraandSpr:2004}Braithwaite,
J. and Spruit, H.C., 2004, Nature, 431, 891.
\bibitem[Braithwaite \& Spruit 2006]{BraandSpr:2006}Braithwaite,
J. and Spruit, H.C., 2006, A\&A, 450, 1097.
\bibitem[Flowers \& Ruderman 1977]{FloandRud:1977}Flowers, E. and Ruderman, M.A., 1977, ApJ, 215, 302.
\bibitem[Frieman \& Rotenberg 1960]{FreandRot:1960}Frieman, E.A. and
Rotenberg, M. 1960, Rev. Mod. Phys. 32, 898.
\bibitem[Geppert \& Rheinhardt 2006]{GepandRhe:2006}Geppert, U. and
Rheinhardt, M., 2006, A\&A, {\mk 456, 639}.
\bibitem[Henrichs et al. 2003]{Henrichsetal:2003}Henrichs, H., Neiner,
C., Geers, V. and de Jong, J., 2003, in ``Magnetism and activity of
the Sun and stars'', J. Arnaud and N. Meunier (eds.), EAS Publ. Series,
9, p.353.
\bibitem[Hyman 1979]{Hyman:1979}Hyman, J., 1979, in R. Vichnevetsky,
R.S. Stepleman (eds.), Adv. in Comp. Meth. for PDEs - III, 313.
\bibitem[Jones et al. 1997]{Jonesetal:1997}Jones, T.B., Washizu, M. and Gans, R., 1997, Jour. Appl. Phys., 82, 883.
\bibitem[Jordan et al. 2005]{Jordanetal:2005}Jordan, S., Werner, K. and
O'Toole, S.J. 2005, A\&A, 432, 273.
\bibitem[Kamchatnov 1982]{Kamchatnov:1982}Kamchatnov, A.M., 1982,
Zh. Eksp. Teor. Fiz., 82, 117.
\bibitem[Kemp et al. 1970]{Kempetal:1970}Kemp, J.C., Swedlund, J.B., Landstreet, J.D. and Angel,
J.R.P. 1970, ApJ, 161, L77.
\bibitem[Landstreet \& Mathys 2000]{LanandMat:2000}Landstreet, J.D. and
Mathys, G., 2000, A\&A, 359, 213.
\bibitem[Markey \& Tayler 1973]{MarandTay:1973}Markey, P. and Tayler,
R.J., 1973, MNRAS, 163, 77.
\bibitem[Markey \& Tayler 1974]{MarandTay:1974}Markey, P. and Tayler,
R.J., 1974, MNRAS, 168, 505.
\bibitem[Mathys et al. 1997]{Mathysetal:1997}Mathys, G., Hubrig, S., Landstreet, J. D., Lanz, T. and Manfroid, J. 1997, A\&AS 123, 353.
\bibitem[Moss 2001]{Moss:2001}Moss, D., 2001, ASP Conf. Series vol. 248, ``Magnetism across the HZ diagram", eds. G. Mathys, S.K. Solanki and D.T. Wickramasinghe, p. 350.
\bibitem[Nordlund \& Galsgaard 1995]{NorandGal:1995}Nordlund, \AA.,
Galsgaard, K., 1995, \\http://www.astro.ku.dk/$\sim$aake/papers/95.ps.gz
\bibitem[Pitts \& Tayler 1985]{PitandTay:1985}Pitts, E. and Tayler,
R.J., 1985, MNRAS, 216, 139.
\bibitem[Roberts 1981]{Roberts:1981}Roberts, P.H., 1981,
Astron. Nach., 302, 65.
\bibitem[Tayler 1973]{Tayler:1973}Tayler, R.J., 1973, MNRAS, 161, 365.
\bibitem[Wright 1973]{Wright:1973}Wright, G.A.E., 1973, MNRAS, 162, 339.

\end{thebibliography}
\end{document}